 \documentclass[
nofootinbib,
reprint,
superscriptaddress, prl,
]{revtex4-1}

\usepackage[caption=false]{subfig}
\usepackage{float}
\usepackage{custom}

\usepackage{amsthm,amssymb}

\begin{document}

\title{Quantum clocks observe classical and quantum time dilation}

\author{Alexander R. H. Smith}
\affiliation{Department of Physics, Saint Anselm College, Manchester, New Hampshire 03102, USA} \email[]{arhsmith@anselm.edu}
\affiliation{Department of Physics and Astronomy, Dartmouth College, Hanover, New Hampshire 03755, USA}

\author{Mehdi Ahmadi}

\affiliation{Department of Mathematics and Computer Science, Santa Clara University, Santa Clara, California 95053, USA}
\email[]{mahmadi@scu.edu\\ }

\date{\today}
 
\begin{abstract}
At the intersection of quantum theory and relativity lies the possibility of a clock experiencing a superposition of proper times. We consider quantum clocks constructed from the internal degrees of relativistic particles that move through curved spacetime. The probability that one clock reads a given proper time conditioned on another clock reading a different proper time is derived. From this conditional probability distribution, it is shown that when the center-of-mass of these clocks move in localized momentum wave packets they observe classical time dilation. We then illustrate a quantum correction to the time dilation observed by a clock moving in a superposition of localized momentum wave packets that has the potential to be observed in experiment. The Helstrom-Holevo lower bound is used to derive a proper time-energy/mass uncertainty relation.
\end{abstract}

\maketitle

\section*{Introduction}
\label{Introduction}
What allowed Einstein to transcend Newton's absolute time was his insistence that time is what is shown by a clock~\cite{einsteinTheoryRelativity1996}:
\begin{quote}   
``[Time is] considered measurable by a clock (ideal periodic process) of negligible spatial extent. The time of an event taking place at a point is then defined as the time shown on the clock simultaneous with the event.''
\end{quote}
Bridgman highlighted the significance of this definition of time~\cite{bridgmanLogicModernPhysics1927}:
\begin{quote}
``Einstein, in seizing on the act of the observer as the essence of the situation, is actually adopting a new point of view as to what the concepts of physics should be, namely, the operational view.''
\end{quote}

Extending the operational view to quantum theory, one is led to define time through measurements of quantum systems serving as clocks~\cite{peresMeasurementTimeQuantum1980}. Such descriptions of quantum clocks have been developed in the context of quantum metrology~\cite{helstromQuantumDetectionEstimation1976,holevoProbabilisticStatisticalAspects1982,braunsteinStatisticalDistanceGeometry1994,braunsteinGeneralizedUncertaintyRelations1996}. In this regard, time observables are identified with positive-operator valued measures (POVMs) that transform covariantly with respect to the group of time translations acting on the employed clock system~\cite{buschOperationalQuantumPhysics,buschQuantumMeasurement2016}. This covariance property ensures that these time observables give the optimal estimate of the time experienced by the clock; that is, they saturate the Cramer-Rao bound~\cite{wisemanQuantumMeasurementControl2010}. Furthermore, covariant time observables allow for a rigorous formulation of the time-energy uncertainty relation~\cite{helstromQuantumDetectionEstimation1976,holevoProbabilisticStatisticalAspects1982,braunsteinStatisticalDistanceGeometry1994,braunsteinGeneralizedUncertaintyRelations1996}, circumvent Pauli's infamous objection to the construction of a time operator~\cite{pauliGeneralPrinciplesQuantum1980,buschTimeObservablesQuantum1994}, and play an important role in relational quantum dynamics~\cite{Brunetti:2009eq,smithQuantizingTimeInteracting2017,loveridgeSymmetryReferenceFrames2018,hoehn2020,hoehnEquivalenceApproachesRelational2020}.
\begin{figure}[t!]
\includegraphics[width= 2.75in]{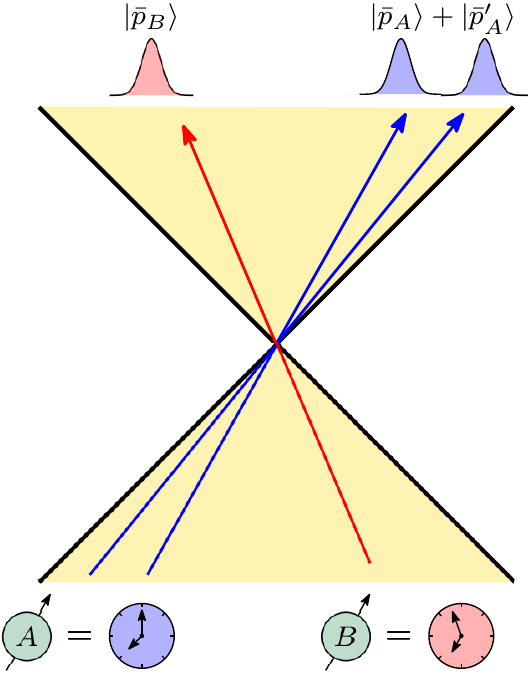}
\caption{{\bf Quantum time dilation.} Two clocks are depicted as moving in Minkowski space. Clock $B$ is moving in a localized momentum wave packet with average momentum~$\bar{p}_B$, while clock $A$ is moving in a superposition of localized momentum wave packets with average momentum $\bar{p}_A$ and~$\bar{p}_A'$. Clock $A$ experiences a quantum contribution to the time dilation it observes relative to clock $B$ due to its nonclassical state of motion.
\label{SpacetimePicture}}
\end{figure}

{Given that clocks are ultimately quantum systems, they too are subject to the superposition principle. In a relativistic context, this leads to the possibility of clocks experiencing a superposition of proper times. Such scenarios have been investigated in the context of relativistic clock interferometry~\cite{zychQuantumInterferometricVisibility2011a}, in which two branches of a matter-wave interferometer experience different proper times on account of either special or general relativistic time dilation~\cite{zychGeneralRelativisticEffects2012a,pikovskiUniversalDecoherenceDue2015,margalitSelfinterferingClockWhich2015a,pangUniversalDecoherenceGravity2016,zychGeneralRelativisticEffects2016, bushevSingleElectronRelativistic2016a,lorianiInterferenceClocksQuantum2019a,rouraGravitationalRedshiftQuantumClock2020a}. Such a setup leads to a signature of matter experiencing a superposition of proper times through a decrease in interferometric visibility.} Other work has focused on quantum variants of the twin-paradox~\cite{vedralSchrodingerCatMeets2008,lindkvistTwinParadoxMacroscopic2014} and exhibiting nonclassical effects in relativistic scenarios~\cite{lockRelativisticQuantumClocks2017,ruizEntanglementQuantumClocks2017,zychGravitationalMassComposite2019a,paigeClassicalNonclassicalTime2020,lockQuantumClassicalEffects2019,khandelwalGeneralRelativisticTime2019,Hoehn:2019,ruizQuantumClocksTemporal2020}.

We introduce a proper time observable defined as a covariant POVM on the internal degrees of freedom of a relativistic particle moving through curved spacetime. This allows us to consider two relativistic quantum clocks, $A$ and $B$, and construct the probability that $A$ reads a particular proper time conditioned on $B$ reading a different proper time. To compute this probability distribution we extend the Page-Wootters approach~\cite{pageEvolutionEvolutionDynamics1983,woottersTimeReplacedQuantum1984} to relational quantum dynamics to the case of a relativistic particle with internal degrees of freedom. We then consider two clocks prepared in localized momentum wave packets and demonstrate that they observe on average classical time dilation in accordance with special relativity. We then illustrate a quantum time dilation effect that occurs when one clock moves in a superposition of two localized momentum wave packets: On average, the proper time of a clock moving in a coherent superposition of momenta is distinct from that of the corresponding classical mixture, see Fig.~\ref{SpacetimePicture}.  We describe the average quantum correction to the classical time dilation observed by such a superposed clock. In addition, our description of proper time as a covariant POVM allows for both proper time and particle mass to be treated as dynamical quantum observables, leading to a time-energy/mass uncertainty~relation.

\section*{Results}
\label{Results}

\subsection*{Page-Wootters description of a relativistic particle with an internal degree of freedom}

In adhering to the operational view espoused earlier, we employ the Page-Wootters formulation of quantum dynamics in which time enters like any other quantum observable. One considers a state $\kket{\Psi} \in \mathcal{H}_{\rm phys}$ of a clock $C$ and a system $S$ that lives in the physical Hilbert space $\mathcal{H}_{\rm phys}$. This Hilbert space is defined as the Cauchy completion of the set of solutions to the constraint equation
\begin{equation}
C_H \kket{\Psi} = \left( H_C + H_S \right) \kket{\Psi} = 0,
\label{Constraint}
\end{equation}
where $H_C \in \mathcal{L}(\mathcal{H}_C)$ and $H_S\in \mathcal{L}(\mathcal{H}_S)$ denote the Hamiltonians of $C$ and $S$. One then associates with $C$ a time observable defined as a POVM
\begin{equation}
    T_C \ce \left\{ E_C(t) \   \forall t \in G \, \bigg| \, I_C=\int_G dt \, E_C(t) \right\},
    \label{MinkowskiTime}
\end{equation}
where $E_C(t) = \ket{t}\!\bra{t}$ is a positive operator on $\mathcal{H}_C$ known as an effect operator, $G$ the group generated by $H_C$, and $\ket{t}$ will be referred to as a \emph{clock state} associated with a measurement of the clock yielding the time $t$. What makes $T_C$ a time observable is that the effect operators transform covariantly with respect to the group generated by $H_C$~\cite{holevoProbabilisticStatisticalAspects1982,braunsteinStatisticalDistanceGeometry1994,braunsteinGeneralizedUncertaintyRelations1996,buschQuantumMeasurement2016}
\begin{align}
    E(t+t') &= U_{C}(t') \, E(t)  \, U_{C}^\dagger (t')
\label{CovarianceCondition}
\end{align}
where $U_C(t) \ce e^{-i H_C t}$. This covariance condition implies that $\ket{t+t'} = U_C(t')\ket{t}$. One then defines a state of $S$ by conditioning $\kket{\Psi}$ on $C$ reading the time $t$
\begin{equation}
    \ket{\psi_S(t)} \ce    \bra{t} \otimes I_S  \kket{\Psi}.
\label{defconditinalstate}
\end{equation}
It then follows from Eqs.~\eqref{Constraint} and \eqref{CovarianceCondition} that
\begin{equation}
    i \frac{d}{dt} \ket{\psi_S(t)} = H_S \ket{\psi_S(t)},
    \label{Schrodinger}
\end{equation}
which describes the evolution of $S$ relative to~$C$.

As described in the ``Methods'' section, a relativistic particle with an internal clock degree of freedom can be described by the Hilbert space $\mathcal{H}_t \otimes \mathcal{H}_{\rm cm} \otimes \mathcal{H}_{\rm clock}$, where $\mathcal{H}_t \simeq L^2(\mathbb{R})$, $\mathcal{H}_{\rm cm} \simeq L^2(\mathbb{R}^3)$, and $\mathcal{H}_{\rm clock}$ are Hilbert spaces associated respectively with the temporal, center-of-mass, and internal clock degrees of freedom of the particle. When the relativistic particle has positive energy, the physical state satisfies Eq.~\eqref{Constraint} with $H_C =P_t$ equal to the momentum operator on $\mathcal{H}_t$ and
\begin{align}
    H_S =  mc^2 \sqrt{ \frac{\mathbf{P}^2}{m^2 c^2  }  + \left(1 + \frac{H_{\rm clock}}{mc^2}\right)^2},
    \label{RelativisticHamiltonian}
\end{align}
where $\mathbf{P}^2 \ce \eta_{ij} P^i P^j$ is the square of the center-of-mass momentum,  $\eta_{ij}$ the Minkowski metric, and $m$ the rest mass of the particle. Equation~\eqref{Schrodinger} then becomes the relativistic Schr\"{o}dinger equation and the state $\ket{\psi_S(t)}$ may be interpreted as the state of the center-of-mass and internal clock of the particle at the time $t$, interpreted as the time of an inertial frame observing the particle with respect to which the center-of-mass degrees of freedom are defined. With this identification, the dynamics implied by the Page-Wootters formalism is  in agreement with previous descriptions of a relativistic particle with internal degrees of freedom~\cite{zychGeneralRelativisticEffects2012a,sonnleitnerMassenergyAnomalousFriction2018,zychGravitationalMassComposite2019a}.

We note that in the ``Methods'' section the above analysis in the Page-Wootters formalism is generalized to the case of a stationary curved spacetime and in Supplementary Note 1 the Klein-Gordon equation is recovered. Further justification of the Page-Wootters formalism is provided in Supplementary Note 2.

\subsection*{Proper time observables}

We now make precise how the internal degrees of freedom of the relativistic particle introduced in the previous section constitute a clock by introducing a proper time observable. We define a clock to be the quadruple:
\begin{equation}
\begin{array}{l}
\includegraphics[height=.3in]{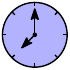} 
\end{array}   =  \, \left\{\mathcal{H}_{\rm clock}, \rho, H_{\rm clock}, T_{\rm clock} \right\},
\end{equation}
the elements of which are the clock Hilbert space~$\mathcal{H}_{\rm clock}$, a fiducial state $\rho$, Hamiltonian $H_{\rm clock}$, and time observable $T_{\rm clock}$. Similar to the definition of $T_C$ above, $T_{\rm clock}$ is defined as a POVM that transforms covariantly with respect to the group action $U_{\rm clock}(\tau) = e^{-iH_{\rm clock} \tau }$. The physical significance of the covariance condition in Eq.~\eqref{CovarianceCondition} is that it implies the time observable satisfies the following two physical properties commonly associated with a clock, which we state in a theorem.

\emph{Theorem 1:} (Desiderata of physical clocks)
Let $T_{\rm clock}$ be a covariant time observable relative to the group generated by $H_{\rm clock}$, $\rho$ be a fiducial state be  such that $\braket{T_{\rm clock}}_{\rho} = 0$, and $\rho(\tau) \ce U_{\rm clock}(\tau) \,  \rho \, U_{\rm clock}^\dagger(\tau)$. The following two physical properties of such a time observable follow:
\begin{enumerate}
\item $T_{\rm clock}$ is an unbiased estimator of the parameter $\tau$ such that $\braket{T_{\rm clock}}_{\rho(\tau)} = \tau$.

\item The variance of the time observable is independent of the parameter $\tau$, i.e.,  ${\braket{(\Delta T_{\rm clock})^2}_{\rho(\tau)}} = \braket{(\Delta T_{\rm clock})^2}_\rho$.
\end{enumerate}

\emph{Proof:} Statements 1 and 2 follow directly from the covariance property of $T_{\rm clock}$; see Supplementary Note 3. 

This theorem justifies interpreting $T_{\rm clock}$ as a time observable: When a time observable is measured on a quantum clock, we expect on average that it estimates the elapsed time $\tau$ unitarily encoded in $\rho(\tau)$. Also the variance of this measurement should be independent of the time $\tau$ being estimated.

Taking this notion of a clock and applying it to the relativistic particle model introduced in the previous section, we may construct a proper time observable that transforms covariantly with respect to the internal clock Hamiltonian $H_{\rm clock}$ of the particle. As explained in the Methods Section, such a Hamiltonian generates a unitary evolution of the internal clock degrees of freedom of the particle, and thus a time observable $T_{\rm clock}$ that transforms covariantly with respect to the group generated by $H_{\rm clock}$ will measure the particle's proper time.

\subsection*{Proper time-energy/mass uncertainty relation}

For an unbiased estimator, like the proper time observable $T_{\rm clock}$ introduced in the previous section, the  Helstrom-Holevo lower bound \cite{helstromQuantumDetectionEstimation1976, holevoProbabilisticStatisticalAspects1982} places the fundamental limit on the variance of the proper time measured by the clock~
\begin{align}
{\braket{(\Delta T_{\rm clock})^2}_{\rho} } \geq \frac{1}{4\braket{(\Delta H_{\rm clock} )^2}_{\rho}},
\label{TimeEnergyUncertanity}
\end{align}
where $\braket{(\Delta H_{\rm clock} )^2}_{\rho}$ is the variance of $H_{\rm clock}$ on the fiducial  state $\rho$. Equation~\eqref{TimeEnergyUncertanity} is a time-energy uncertainty relation between the proper time estimated by $T_{\rm clock}$ and a measurement of the clock's energy~$H_{\rm clock}$. Now consider the related mass observable defined by the self-adjoint operator  $M_{\rm clock} \ce m + H_{\rm clock}/c^2$ (e.g.,~\cite{zychGeneralRelativisticEffects2012a}). From Eq.~\eqref{TimeEnergyUncertanity}, an uncertainty relation between this mass observable and proper time follows
\begin{align}
\Delta {M_{\rm clock}} \Delta {T_{\rm clock}} \geq \frac{1}{2c^2}, 
\label{timeEnergy}
\end{align}
where $\Delta A \ce  \braket{(\Delta A)^2}_{\rho}^{1/2}$ denotes the standard deviation of the observable $A$. This inequality gives the ultimate bound on the precision of any measurement of proper time. 

The time-energy/mass inequality above can be saturated using the optimal proper time observable provided that the effect operators $E_{\rm clock}(\tau)$ defining $T_{\rm clock}$ are proportional to `projection' operators
\begin{align}
E_{\rm clock}(\tau) = \mu \ket{\tau}\!\bra{\tau},
\label{projectionCondition}
\end{align}
for $\mu\in \mathbb{R}$ such that $\int_G d\tau \, E_{\rm clock}( \tau) = I_{\rm clock}$, where $H_{\rm clock}$ is the generator of proper time translations, and $\ket{\tau}$ is the clock state corresponding to the proper time~$\tau$. Here the motivation is that measurements not described by one-dimensional projectors have less resolution~\cite{braunsteinStatisticalDistanceGeometry1994}, however, note that the clock states $\ket{\tau}$ are not necessarily orthogonal, $\braket{\tau | \tau '} \neq 0$.

It turns out that covariant observables satisfying  Eq.~\eqref{projectionCondition} constitute an optimal measurement to estimate the parameter $\tau$  unitarily encoded in the state $\rho(\tau) \ce U_{\rm clock}(\tau) \rho U_{\rm clock}(\tau)^\dagger$, provided that the fiducial state is pure $\rho = \ket{\psi_{\rm clock}}\! \bra{\psi_{\rm clock}}$  and
\begin{equation}
\ket{\psi_{\rm clock}} = \int_G d\tau \, \abs{\psi_{\rm clock}(\tau)}  e^{i \tau \braket{H_{\rm clock}}_{ \rho}} \ket{\tau},
\label{intialClockState}
\end{equation}
where $\psi_{\rm clock}(\tau) \ce \sqrt{\mu} \braket{\tau | \psi_{\rm clock}} $ and $\abs{\psi_{\rm clock}(\tau)}$ is a real function of $\tau$ such that \mbox{$\braket{T_{\rm clock}}_{\rho} = 0$} \cite{braunsteinStatisticalDistanceGeometry1994}. Such a proper time observable $T_{\rm clock}$ is optimal in the sense that it maximizes the so-called Fisher information~\cite{wisemanQuantumMeasurementControl2010}, which quantifies how well two slightly different values of proper time can be distinguished given a particular quantum measurement. For the effect operators $E_{\rm clock}(\tau)$ and the fiducial state in Eq. \eqref{intialClockState}, we have
\begin{align}
F[\tau; \rho(\tau)] = 4 \braket{(\Delta H_{\rm clock} )^2}_{\rho}.
\label{CRoptimal}
\end{align}
The covariance condition, $\ket{\tau+ \tau'} = U_{\rm clock}(\tau') \ket{\tau}$, ensures that the Fisher information is independent of $\tau$.

Let us point out a connection between our above construction of a proper time observable and quantum speed limits. From the fact that  $\sqrt{F\left[\tau;\rho(\tau)\right]}/2 \leq  \Delta H $~\cite{MT2}, together with Eq.~\eqref{CRoptimal}, we can conclude that the covariant proper time observable in fact saturates the so-called Mandelstam and Tamm inequality. That is
\begin{align}
\tau_{\perp} = \Delta T_{\rm clock}  =\frac{\pi}{\sqrt{F\left[\tau;\rho(\tau)\right]}} = \frac{\pi}{2 \Delta M_{\rm clock}},
\end{align}
where $\tau_{\perp}$ is the time that passes before the initial state of a system evolves under the Hamiltonian $H_{\rm clock}$ into an orthogonal state.

{We remark that in this construction both proper time and mass are treated as genuine quantum observables; the former as a covariant POVM $T_{\rm clock}$ and the latter as a self-adjoint operator $M_{\rm clock}$. Such a formulation of proper time and mass in the regime of relativistic quantum mechanics has been argued as necessary by Greenberger~\cite{greenbergerConceptualProblemsRelated2010,Greenberger:2018}.}

\subsection*{Classical and quantum time dilation}

Let us now consider two relativistic particles $A$ and $B$, each with an internal degree of freedom serving as a clock, $\{\mathcal{H}_A^{\rm clock}, \ket{\psi_A^{\rm clock}}, H_A^{\rm clock}, T_A\}$ and $ \{\mathcal{H}_B^{\rm clock}, \ket{\psi_B^{\rm clock}}, H_B^{\rm clock}, T_B\}$. Suppose these clocks move through Minkowski space and are described by the physical state $\kket{\Psi}$, which satisfies two copies of Eq.~\eqref{Constraint}, one for each clock, as detailed in the Methods Section. To probe time dilation effects between these clocks we  consider  the probability that clock $A$ reads the proper time $\tau_A$ conditioned on clock $B$ reading the proper time $\tau_B$. This conditional probability is evaluated using the physical state $\kket{\Psi}$ and the Born rule as follows
\begin{align}
\prob \left[T_A = \tau_A \ | \ T_B = \tau_B \right] &= \frac{\prob \left[T_A = \tau_A \ \& \ T_B = \tau_B  \right] }{\prob \left[ T_B = \tau_B  \right]} \nn \\
&= \frac{\bbra{\Psi} E_{A}(\tau_A)  E_{B}(\tau_B)  \kket{\Psi} }{\bbra{\Psi} E_{B}(\tau_B) \kket{\Psi}}.
\label{ConditionalProbabilityDis} 
\end{align}
To evaluate this probability distribution note that the clock states defined below Eq.~\eqref{MinkowskiTime} form a basis dense in $\mathcal{H}_t$, and thus a physical state may be expanded as
\begin{align}
\kket{\Psi} &=  \int dt\, \ket{t}\ket{\psi_S(t)} \nn \\
&=  \int dt\, \ket{t} \bigotimes_{n \in \{A,B\}}  U_n(t) \ket{\psi^{\rm cm}_{n}} \ket{\psi^{\rm clock}_n} ,
\label{PhysicalStateAB}
\end{align}
where $U_n(t) \ce e^{-iH_n t}$, $H_n$ is the Hamiltonian given in Eq.~\eqref{RelativisticHamiltonian}, and in writing Eq.~\eqref{PhysicalStateAB} we have supposed that the conditional state at $t=0$ is unentangled, $\ket{\psi_S(0)} = \ket{\psi_{A}} \ket{\psi_{B}}$. Further suppose that the center-of-mass and internal clock degrees of freedom of both particles are unentangled, $\ket{\psi_{n}} = \ket{\psi^{\rm cm}_{n}} \ket{\psi^{\rm clock}_n}$, where $\ket{\psi^{\rm cm}_{n}} \in \mathcal{H}^{\rm cm}_{n} $ is the initial state of the center-of-mass of the particle. Suppose that $\mathcal{H}_{n}^{\rm clock} \simeq L^2(\mathbb{R})$, so that we may consider an ideal clock such that $H_n^{\rm clock} = P_n$ and $T_n$ are the momentum and position operators on $\mathcal{H}_n^{\rm clock}$. Such clocks represent a commonly used idealization in which the time observable is sharp, that is, the clock states are orthogonal $\braket{\tau_i|\tau'_i} = \delta(\tau-\tau')$ and so outcomes of different clock measurements are perfectly distinguishable. Note that $\left[ T_{n} , H^{\rm clock}_{n} \right]  = i$, from which it follows that the effect operators satisfy the covariance relation $\ket{\tau + \tau'} = U_{\rm clock}(\tau') \ket{\tau}$. We employ such clocks for their mathematical simplicity in illustrating the quantum time dilation effect, however we stress that for any covariant time observable, on account of Eq.~\eqref{ConditionalProbabilityDis}, a quantum time dilation effect is expected~(e.g., \cite{grochowskiQuantumTimeDilation2020}).

By substituting Eq.~\eqref{PhysicalStateAB} into Eq.~\eqref{ConditionalProbabilityDis}, the probability that $A$ reads $\tau_A$ conditioned on $B$ reading $\tau_B$ can be evaluated to leading relativistic order in the center-of-mass momentum $\braket{\mathbf{P}_n/mc}$ and internal clock energy $\braket{H^{\rm cm}_{n}/mc^2}$
\begin{align}
&\prob \left[T_A = \tau_A \ | \ T_B = \tau_B \right] \nn \\
&= \frac{e^{- \frac{(\tau_A-\tau_B)^2}{2 \sigma^2}}}{\sqrt{2 \pi} \sigma}  \left[ 1 +   \frac{\braket{H^{\rm cm}_{A}}- \braket{H^{\rm cm}_{B}}}{2 mc^2}  \left(1 - \frac{ \tau_A^2 - \tau_B^2}{ \sigma^2}  \right)\right],
\label{ConditionalProbabilityDistirbutionMinkowski}
\end{align}
where $\braket{H_n^{\rm cm }} \ce \bra{\psi^{\rm cm}_{n}}  \mathbf{P}^2_n \ket{\psi^{\rm cm}_{n}}/2m$ is the average non-relativistic kinetic energy of the $n$th particle and we have assumed the fiducial states of the clocks $\ket{\psi_{n}^{\rm clock}}$ to be Gaussian wave packets centered at $\tau=0$ and have a width $\sigma$ in the clock state (i.e., position) basis. As described by this probability distribution, the average proper time read by clock $A$ conditioned on clock $B$ indicating the  time $\tau_B$ is
\begin{align}
\avg{T_A} 
&= \left( 1 -   \frac{\braket{H^{\rm cm}_{A}}  - \braket{H^{\rm cm}_{B}}}{mc^2}   \right)\tau_B ,
\label{averageEffect}
\end{align}
and the variance in such a measurement is
\begin{align}
{\braket{(\Delta T_A)^2} }
&= \left( 1 -   \frac{\braket{H^{\rm cm}_{A}}  - \braket{H^{\rm cm}_{B}}}{mc^2} \right) \sigma^2.
\label{varianceEffect}
\end{align}
As might have been anticipated, the variance in a measurement of $T_A$ is proportional to $\sigma^2$, which quantifies the spread in the fiducial clock state.

Now suppose that the center-of-mass of both clocks are prepared in a Gaussian state localized around an average momentum $\bar{\mathbf{p}}_n$ with spread~$\Delta_n >0$,
\begin{align}
\ket{\psi^{\rm cm}_{n}} &= \frac{1}{\pi ^{1/4} \sqrt{\Delta_n}} \int d\mathbf{p} \, e^{-\frac{(\mathbf{p}-\bar{\mathbf{p}}_n)^2}{ 2\Delta_n^2}}  \ket{\mathbf{p}_n} 
=: \ket{\bar{\mathbf{p}}_n},
\label{GaussCMState}
\end{align}
for which $
\braket{H_n^{\rm cm }} =\frac{\bar{\mathbf{p}}_n^2}{2m} + \frac{\Delta_n^2}{4m}$. It follows that the observed average time dilation between two such clocks is 
\begin{align}
\braket{T_A}
&= \left[1 - \frac{ \bar{\mathbf{p}}_A^2 - \bar{\mathbf{p}}_B^2 +  \tfrac{1}{2}(\Delta_A^2 - \Delta_B^2)  }{2m^2c^2} \right] \tau_B . \label{averageProperTime}
\end{align}
If instead the two clocks were classical, moving with momenta $\bar{\mathbf{p}}_A$ and $\bar{\mathbf{p}}_B$ corresponding to the average velocity of the momentum wave packets of the clocks just considered, then to leading relativistic order the proper time $\tau_A$ read by $A$ given that $B$ reads the proper time $\tau_B$~is
\begin{equation}
\tau_A = \frac{\gamma_B}{\gamma_A} \tau_B = \left[ 1 - \frac{ \bar{\mathbf{p}}_A^2 - \bar{\mathbf{p}}_B^2} {2m^2c^2}\right]  \tau_B , \label{classicaltimedilation}
\end{equation}
where $\gamma_n \ce \sqrt{1+\bar{\mathbf{p}}_n^2/m^2c^2}$. Therefore, upon comparison with Eq.~\eqref{averageProperTime} and supposing that $\Delta_A = \Delta_B$, quantum clocks whose center-of-mass are prepared in Gaussian wave packets localized around a particular momentum agree on average with classical time dilation described by special relativity.

It is natural to now ask: Does a quantum contribution to the time dilation observed by these clocks arise if the center-of-mass of one of the clocks moves in a superposition of momenta? To answer this question, suppose that the center-of-mass state of $A$ begins in a superposition of two Gaussian wave packets with average momenta $\bar{\mathbf{p}}_{A}$ and~$\bar{\mathbf{p}}_{A}'$,
\begin{align}
\ket{\psi^{\rm cm}_A}  \sim \cos \theta \ket{\bar{\mathbf{p}}_{A}} + e^{i\phi}   \sin \theta \ket{\bar{\mathbf{p}}_{A}'},
\label{nonclassical}
\end{align}
where $\theta \in [0,\pi/2)$, $\phi \in [0,\pi]$, and $\ket{\bar{\mathbf{p}}_{A}}$ and $\ket{\bar{\mathbf{p}}'_{A}}$ are defined in Eq.~\eqref{GaussCMState}. Further, suppose that the center-of-mass degree of freedom of clock $B$ is again prepared in a Gaussian wave packet with average momentum $\bar{\mathbf{p}}_B$ as in Eq.~\eqref{GaussCMState}. Using Eq.~\eqref{averageEffect}, the average time read by $A$ conditioned on $B$ reading $\tau_B$~is
\begin{align}
    \avg{T_A} = \left( \gamma_{\rm C}^{-1} + \gamma_{\rm Q}^{-1} \right) \tau_B,
\end{align}
where
\begin{align}
    \gamma_{\rm C}^{-1} \ce 1- \frac{ \bar{\mathbf{p}}_A^2 \cos^2 \theta +  \bar{\mathbf{p}}_A'^2 \sin^2 \theta - \bar{\mathbf{p}}_B^2}{2m^2 c^2 } - \frac{\Delta_A^2- \Delta_B^2}{4m^2 c^2 },
    \label{gammaC}
\end{align}
leads to the classical time dilation expected by a clock moving in a statistical mixture of momenta $\bar{\mathbf{p}}_A$ and $\bar{\mathbf{p}}_A'$ with probabilities $\cos^2 \theta$ and $\sin^2 \theta$, and\\
\begin{equation}
    \gamma_{\rm Q}^{-1} \ce \frac{ \cos \phi \sin 2 \theta \! \left[ \left(\bar{ \mathbf p}_A'-\bar{ \mathbf p}_A\right)^2-2 \!\left(\bar{\mathbf p}_A'^2-\bar{\mathbf p}_A^2\right) \! \cos 2 \theta \right] }{8 m^2 c^2 \left(\cos \phi \sin 2 \theta +\exp \left[\frac{\left( \bar{\mathbf p}_A' - \bar{\mathbf p}_A \right)^2}{4 \Delta_A^2} \right] \right) },
    \label{gammaQ}
\end{equation}
which quantifies the quantum contribution to the time dilation between $A$ and $B$. As expected, if either \mbox{$\theta \in \{0, \tfrac{\pi}{2} \}$} or \mbox{$\bar{\mathbf{p}}_A=\bar{\mathbf{p}}_A' $}, then the quantum contribution vanishes,  \mbox{$\gamma_{\rm Q}^{-1} =0$}. This is expected given that in these cases the center-of-mass of the clock particle is no longer a superposition of momentum wave packets; see Eq.~\eqref{nonclassical}. From Eq.~\eqref{gammaC} it is clear that choosing $\Delta_A = \Delta_B$ and $\bar{\mathbf{p}}_B^2 = \bar{\mathbf{p}}_{A}^2 \cos^2 \theta + \bar{\mathbf{p}}_{A}'^2 \sin^2 \theta$ results in $\gamma_{\rm C}^{-1} =0$, and thus any observed time dilation between clocks $A$ and $B$ would be a result of the quantum contribution~$\gamma_{\rm Q}^{-1}$.

\begin{figure}[t!]
\includegraphics[width=\linewidth]{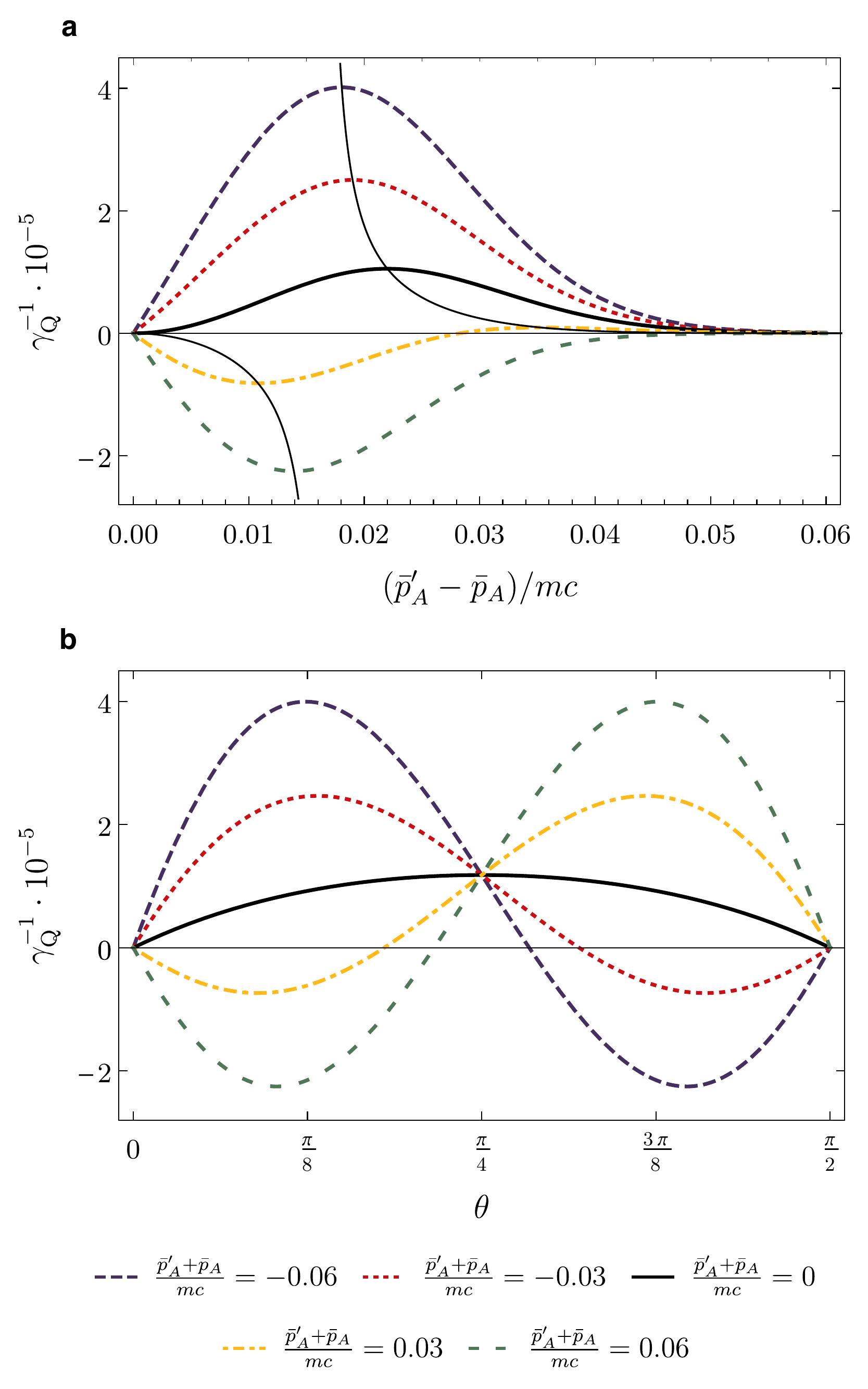}
\caption{{\bf Magnitude of quantum time dilation.} The strength of the quantum time dilation effect $\gamma^{-1}_{\rm Q}$ is plotted in {\sf a } as a function of the momenta difference $(\bar{p}_A'-\bar{p}_A)/mc$, where $\bar{\mathbf{p}}_A = (\bar{p}_A,0,0)$ and $\bar{\mathbf{p}}_A' = (\bar{p}_A',0,0)$ denote the average momentum of the wave packets comprising the superposition state in Eq.~\eqref{nonclassical} for $\theta = \pi/8$, and in {\sf b} as a function of superposition weight $\theta$ for $(\bar{p}_A'-\bar{p}_A)/mc = 0.17$. Different values of average total momentum $(\bar{p}_A'+\bar{p}_A)/mc$ are shown and $\Delta_A/mc = 0.01$ in all cases. The thin black line in plot~{\sf  a} traces the trajectory of the optimal momentum difference $\bar{p}_{\rm opt}$ for different total momentum $\left( \bar{p}_A'+\bar{p}_A\right)/mc$.  }

\label{NonClassicalStates}
\end{figure}

To illustrate the behaviour of quantum  time dilation stemming from the nonclassicality of the center-of-mass state of clock $A$, the quantity $\gamma_{\rm Q}^{-1}$ is plotted in Fig.~\ref{NonClassicalStates}. For simplicity the one-dimensional case is exhibited by supposing $\bar{\mathbf{p}}_A = (\bar{p}_A ,0,0)$ and $\bar{\mathbf{p}}_A' = (\bar{p}_A' ,0,0)$ with $\bar{p}_A' > \bar{p}_A$. Figure~\ref{NonClassicalStates} (top) shows the behaviour of $\gamma_{\rm Q}^{-1}$ as a function of the difference $\left( \bar{p}_A' - \bar{p}_A \right)/mc$ in the average momentum of each wave packet comprising the momentum superposition in Eq.~\eqref{nonclassical} for different values of their total momentum $\left( \bar{p}_A' + \bar{p}_A\right)/mc$. It is seen that quantum time dilation can be either positive or negative, corresponding to increasing or decreasing the total time dilation experienced by the clock compared to an equivalent clock moving in a classical mixture of the same momenta wave packets. Further, there is an optimal difference in the average momentum of the two wave packets $p_{\rm opt}$; as the total average momentum of the wave packets $\left| \bar{p}_A' + \bar{p}_A\right|/mc$ increases, the magnitude of $\gamma_{\rm Q}^{-1}$ and $p_{\rm opt}$ increase.

Figure~\ref{NonClassicalStates} (bottom) is a plot of $\gamma_{\rm Q}^{-1}$ as a function of $\theta$ quantifying the weight of each momentum wave packet comprising the superposition in Eq.~\eqref{nonclassical} for a fixed value of the difference in average momentum of each wave packet $\left( \bar{p}_A' - \bar{p}_A \right)/mc$. It is observed that when $\left( \bar{p}_A' + \bar{p}_A \right)/mc = 0$, $\gamma_{\rm Q}^{-1}$ is positive for all values of $\theta$ and reaches its maximum value at $\theta = \pi/4$. As the total average momentum increases, $\gamma_{\rm Q}^{-1}$ decreases for $0< \theta < \pi/4$ and increases for  $\pi/4 < \theta < \pi/2$ with the largest negative value at  $\theta \approx\pi/8$ and largest positive value at $\theta \approx 3\pi/8$.

\subsection*{Quantum time dilation in experiment}
\label{Quantum time dilation in experiment}

Consider the two clock particles to be ${^{87}}$Rb atoms, which have a mass of $m = 1.4 \times 10^{-25}$kg and atomic radius of $r = \hbar/\Delta_A =  2.5 \times 10^{-10}$m. Suppose these clock particles can be prepared in a superposition of momentum wave packets such that $\left( \bar{p}_A' + \bar{p}_A \right)/mc = \left( \bar{p}_A' - \bar{p}_A \right)/mc=  6.7 \times 10^{-8}$, corresponding to each branch of the momentum superposition moving at average velocities of $\bar{v}_A=5$\,m\,s$^{-1}$ and $\bar{v}_A' =15$\,m\,s$^{-1}$. We note that (classical) special relativistic time dilation has been observed with atomic clocks moving at these velocities \cite{reinhardtTestRelativisticTime2007, chouOpticalClocksRelativity2010}, and perhaps the momentum superposition can be prepared by a momentum beam splitter realized using coherent momentum exchange between atoms and light~\cite{Berman:1997,cladeLargeMomentumBeamsplitter2009}. Supposing that $\theta = 3 \pi/4$ and $\phi =0$ results in $\gamma_{\rm Q}^{-1} \approx  10^{-15}$. Assuming that the resolution of the clock formed by the internal degrees of freedom of the ${^{87}}$Rb atoms is $10^{-14}$s, corresponding to the resolution of ${^{87}}$Rb atomic clocks~\cite{camparoRubidiumAtomicClock2007}, it is seen from Eq.~\eqref{averageEffect} that the coherence time of the momentum superposition must be on the order of 10s to observe a quantum time dilation effect. We note that the required coherence time is comparable to coherence times of the superpositions created in the experiments of Kasevich \emph{et al.}~\cite{kovachyQuantumSuperpositionHalfmetre2015}, which are on the order of seconds. 

Concretely, one might imagine observing quantum time dilation in a spectroscopic experiment using the width of an emission line, which is inversely proportional to the lifetime of the associated excited state, as a quantum clock. Indeed, it has recently been shown that the lifetime of an excited hydrogen-like atom moving in a superposition of relativistic momenta experiences quantum time dilation in accordance with Eq.~\eqref{gammaQ}~\cite{grochowskiQuantumTimeDilation2020}. Alternatively, Bushev et al.~\cite{bushevSingleElectronRelativistic2016a} have proposed to use the spin precession of a single electron in a Penning trap as a clock to observe nonclassical relativistic time dilation effects, and it is conceivable that such a clock might be able to witness the quantum time dilation effect discussed here. Similar remarks apply to the ion trap atomic clock discussed in~\cite{paigeClassicalNonclassicalTime2020}.

\section*{Discussion}
\label{Discussion}

We considered the internal degrees of freedom of relativistic particles to function as clocks that track their proper time. In doing so we constructed an optimal covariant proper time observable which gives an unbiased estimate of the clock's proper time. It was shown that the Helstrom-Holevo lower bound \cite{helstromQuantumDetectionEstimation1976, holevoProbabilisticStatisticalAspects1982} implies a time-energy uncertainty relation between the proper time read by such a clock and a measurement of its energy. From this relation, we {derived} an uncertainty relation between proper time and mass, which {provided} the ultimate bound on the precision of any measurement of proper time. This yielded a consistent treatment of mass and proper time as quantum observables related by an uncertainty relation, resolving past issues with such an approach~\cite{greenbergerConceptualProblemsRelated2010,Greenberger:2018}. The approach adopted here differs in that we construct a proper time observable $T_{\rm clock}$ in terms of a covariant POVM rather than a self-adjoint operator. Using the standard Born rule, the conditional probability distribution that one such clock reads the proper time $\tau_A$ conditioned on another clock reading the proper time $\tau_B$ was derived in Eq.~\eqref{ConditionalProbabilityDis}.

We then specialized to two such clock particles moving through Minkowski space and evaluated the leading-order relativistic correction to this conditional probability distribution. It was shown that on average these quantum clocks measure a time dilation consistent with special relativity when the state of their center-of-mass is localized in momentum space. However, when the state of their center-of-mass is in a superposition of such localized momentum states, we demonstrated that a quantum time dilation effect occurs. We exhibited how this quantum time dilation depends on the parameters defining the momentum superposition and gave an order of magnitude estimate for the size of this effect. We conclude that quantum time dilation may be observable with present day technology, but note that the experimental feasibility of observing this effect remains to be explored.

It should be noted that the conditional probability distribution in Eq.~\eqref{ConditionalProbabilityDis} associated with clocks reading different times was a nonperturbative expression for clocks in arbitrary nonclassical states in a curved spacetime. It thus remains to investigate the effect of other nonclassical features of the clock particles such as shared entanglement among the clocks and spatial superpositions. In regard to the latter, it will be interesting to recover previous relativistic time dilation effects in quantum systems related to particles prepared in spatial superpositions and each branch in the superposition experiencing a different proper time due to gravitational time dilation~\cite{zychGeneralRelativisticEffects2012a,pikovskiUniversalDecoherenceDue2015, zychQuantumInterferometricVisibility2011a, Anastopoulos:2018, khandelwalGeneralRelativisticTime2019}. We emphasize that the quantum time dilation effect described here differs from these results in that it is a consequence of a momentum superposition rather than gravitational time dilation. Nonetheless, it will be interesting to examine such gravitational time dilation effects in the framework developed above and make connections with previous literature on quantum aspects of the equivalence principle \cite{violaTestingEquivalencePrinciple1997,zychQuantumFormulationEinstein2018,hardyImplementationQuantumEquivalence2019}. We also note that while we exhibited the quantum time dilation effect for a specific clock model, based on the preceding analysis in terms covariant time observables it is expected that any clock will witness quantum time dilation. Given this, it will be fruitful to examine our results in relation to other models of quantum clocks that have been considered~\cite{saleckerQuantumLimitationsMeasurement1958,peresMeasurementTimeQuantum1980,buzekOptimalQuantumClocks1999,erkerAutonomousQuantumClocks2017,woodsAutonomousQuantumMachines2018,paigeClassicalNonclassicalTime2020} and establish whether quantum time dilation is universally, affecting all clocks in the same way, like its classical counterpart.

Another avenue of exploration is the construction of  relativistic quantum reference frames from  the relativistic clock particles considered here~\cite{bartlettRelativisticallyInvariantQuantum2005,bartlettReferenceFramesSuperselection2007, bartlettQuantumCommunicationUsing2009, ahmadiCommunicationInertialObservers2015, loveridgeSymmetryReferenceFrames2018}. In particular, one might define relational coordinates with respect to a reference particle and examine the corresponding relational quantum theory and the possibility of changing between different reference frames \cite{poulinToyModelRelational2006, angeloPhysicsQuantumReference2011, palmerChangingQuantumReference2014, Safranek2015, smithQuantumReferenceFrames2016, smithCommunicatingSharedReference2018,giacominiRelativisticQuantumReference2018, giacominiQuantumMechanicsCovariance2019}. Related is the perspective-neutral interpretation of the Hamiltonian constraint in terms of which a formalism for changing clock reference systems has recently been developed~\cite{vanrietveldeChangePerspectiveSwitching2018, Hoehn:2019, hoehnHowSwitchRelational2018, vanrietveldeSwitchingQuantumReference2018}.

\section*{Methods}

\subsection*{\small Constraint description of relativistic particles with internal degrees of freedom}
\label{Relativistic quantum clock particles}
\begin{figure}[t]
\includegraphics[width= 3.1in]{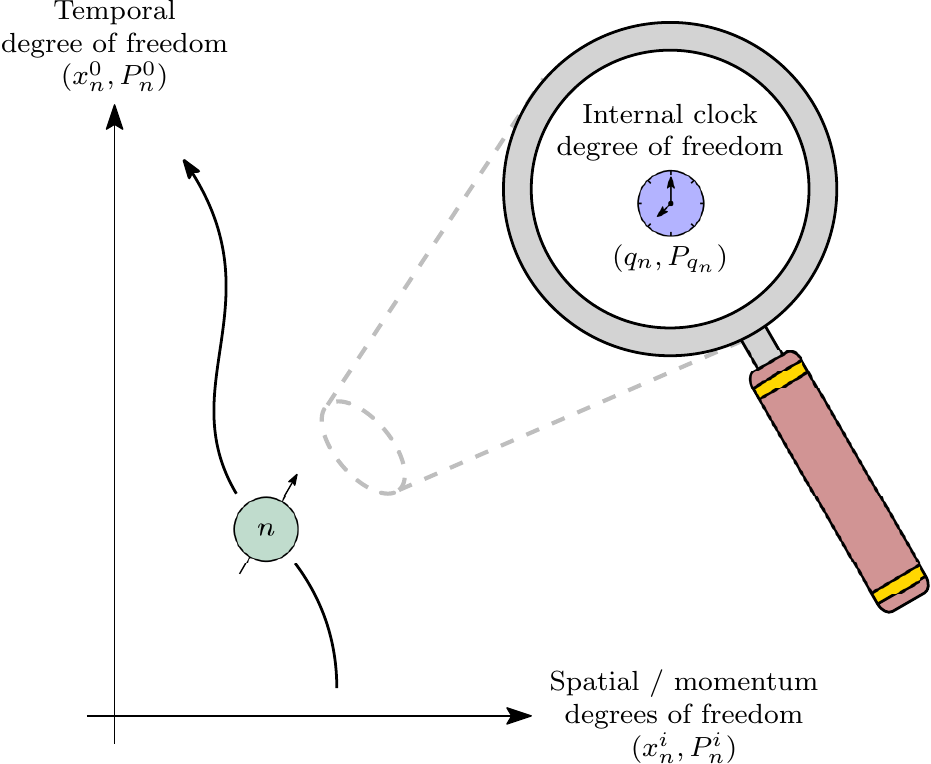}
\caption{ {\bf Degrees of freedom of $n$th particle.} The temporal degrees of freedom $(x^0_n, P^0_n)$ and center-of-mass degrees of freedom $(x^i_n, P^i_n)$ of the $n$th relativistic particle are depicted. This particle carries an internal degree of freedom $(q_{n}, P_{q_n})$ that is used to construct a clock which measures the $n$th particle's proper time.
}
\label{picture1}
\end{figure}

{\small  We present a Hamiltonian constraint formulation of $N$ relativistic particles with internal degrees of freedom. A complementary approach has been taken in~\cite{zychGravitationalMassComposite2019a}.

 Consider a system of $N$ free relativistic particles each carrying a set of internal degrees of freedom, labeled collectively by the configuration variables $q_n$ and their conjugate momentum $P_{q_n}$ ($n=1,\ldots, N$), and suppose these particles are moving through a curved spacetime described by the metric $g_{\mu \nu}$. The action describing such a system is $S  =   \sum_{n} \int d\tau_n \, L_n (\tau_n)$, where
\begin{align}
L_n(\tau_n) \ce  -m_nc^2 +  P_{q_n} \frac{dq_n}{d\tau_n} - H^{\rm clock}_n,
\label{Lagrangian1}
\end{align}
is the Lagrangian associated with the $n$th particle, $\tau_n$ and $m_n$ denote respectively this particle's proper time and rest mass, and $H^{\rm clock}_n =  H^{\rm clock}_n \smash{(q_n, P_{q_n})}$ is the Hamiltonian governing its internal degrees of freedom. We use these internal degrees of freedom as a clock tracking the $n$th  particle's proper time. Note that Eq.~\eqref{Lagrangian1} specifies that $H^{\rm clock}_n$ generates an evolution of the internal degrees of freedom of the $n$th particle with respect to its proper time. Let $x_n^{\mu}$ denote the spacetime position of the $n$th particle's center-of-mass relative to an inertial observer, see Fig.~\ref{picture1}.

The differential proper time $d\tau_n$ along the $n$th particle's world line $x^{\mu}_n(t_n)$, parametrized in terms of an arbitrary parameter $t_n$, is 
\begin{align}
d\tau_n =  \sqrt{-g_{\mu \nu}^{\,}\smash{\dot{x}_n^\mu \dot{x}_n^\nu}/c^2 \vphantom{X}} \, dt_n = \sqrt{-\dot{x}^2_n} \,dt_n,
\end{align}
where the over dot denotes differentiation with respect to $t_n$ and we have used the dimensionless shorthand $\dot{x}_{n}^2 \ce g_{\mu \nu}\dot{x}_{n}^{\mu} \dot{x}_n^\nu / c^2$. In terms of the parameters $t_n$ the action takes the form
\begin{align}
S &=   \sum_{n} \int dt_n  \, \sqrt{-\dot{x}^2_n} \, L_n (t_n). \label{action2}
\end{align}
The action in Eq.~\eqref{action2} is invariant under changes of the world line parameters $t_n$, as long as there is a one-to-one correspondence between $t_n$ and $\tau_n$. This invariance allows for the action to instead be parameterized in terms of a single parameter $t$, which is connected to the $n$th particle's proper time through a monotonically  increasing function $  f_n(\tau_n) \ce  t$. Expressed in terms of the single parameter $t$~\cite{marneliusLagrangianHamiltonianFormulation1974}, the action in Eq.~\eqref{action2} is \mbox{$S = \int dt \, L(t)$}, where
\begin{align}
L(t) \ce \sum_n \sqrt{-\dot{x}_{n}^2  } \left( -m_n c^2 +   \frac{P_{q_n} \dot{q}_n}{\sqrt{-\dot{x}_{n}^2 }} - H^{\rm clock}_n \right). \label{SuperLagrangian}
\end{align}
This Lagrangian treats the temporal, spatial, and internal degrees of freedom as dynamical variables on equal footing described by an extended phase space interpreted as the description of the particles with respect to an inertial observer.}

{\small The Hamiltonian associated with $L(t)$ is constructed by a Legendre transform of Eq.~\eqref{SuperLagrangian}, which yields
\begin{align}
H &\ce  \sum_{n} \left[ g_{\mu \nu} P^\mu_n \dot{x}_{n}^\nu +  P_{q_n} \dot{q}_n \right] - L(t) \nn \\
 &= \sum_n \left[ g_{\mu \nu} P^\mu_n \dot{x}_{n}^\nu + \sqrt{-\dot{x}_{n}^2 } \left( m_nc^2 + H^{\rm clock}_n \right)\right], \label{Hamiltonian1}
\end{align}
where $P^\mu_n$ is the momentum conjugate to the $n$th particle's spacetime position $x_n^\mu$  defined as
\begin{align}
P_n^\mu \ce g^{\mu\nu} \frac{\partial L(t)}{\partial \dot{x}_n^\nu } = \frac{\dot{x}_n^\mu}{ \sqrt{-\dot{x}_{n}^2}} M_n, \label{conjugateP}
\end{align}
where we have defined the mass function $M_n \ce m_n  + H^{\rm clock}_n/c^2$, comprised of the non-dynamic rest mass $m_n$ and the dynamic mass $H^{\rm clock}_n/c^2$ implied by mass-energy equivalence.
Upon substituting Eq.~\eqref{conjugateP} into the Hamiltonian $H$, we see that each term in Eq.~\eqref{Hamiltonian1} is constrained to vanish
\begin{align}
H_n 
 & =   \frac{g_{\mu \nu} \dot{x}_n^\mu \dot{x}_n^\nu}{\sqrt{-\dot{x}_{n}^2}} M_n + \sqrt{-\dot{x}_{n}^2 } \left( m_n c^2 + H^{\rm clock}_n\right) \nn \\
 & = \frac{\dot{x}_{n}^2}{\sqrt{-\dot{x}_{n}^2}}\left [  \left( m_n c^2 + H^{\rm clock}_n \right) - \left( m_n c^2 + H^{\rm clock}_n\right) \right] \nn \\
 &\approx 0, \label{constraint1}
\end{align}
where $\approx$ means $H_n$ vanishes as a constraint~\cite{diracLecturesQuantumMechanics1964}. Furthermore,
using Eq.~\eqref{conjugateP}, the $N$ constraints in Eq.~\eqref{constraint1} can be expressed as
\begin{align}
C_{H_n}  \ce g_{\mu\nu}P_n^\mu P_n^\nu c^2 +  M_n^2 c^4 \approx 0. \label{constraint2}
\end{align}
 This is a collection of primary first class constraints, which are  quadratic in the particles' momentum and are a manifestation of the Lorentz invariance of the action defined by Eq.~\eqref{Lagrangian1}. 

Similar to~\cite{hoehnHowSwitchRelational2018,vanrietveldeChangePerspectiveSwitching2018}, each of these constraints may be factorized as $C_{H_n} = C_n^+ C_n^-$~, where $C_n^\pm$ is defined as
\begin{align}
C_n^\pm  \ce \sqrt{g_{00}}  (P_n)_0 \pm h_n,
\label{constraint3}
\end{align}
and
\begin{align}
h_n 
&\ce    \sqrt{ g_{ij} P_n^i P_n^j c^2 +  M_n^2 c^4 }. \label{defofh}
\end{align}
In Eq.~\eqref{constraint3} we have assumed the time-space components of the metric vanish, $g_{0i}=0$. Such an assumption is not necessary, however to illustrate the quantum time dilation effect we  will specialize to clocks in Minkowski space for which this is the case.

By construction, the momenta conjugate to the $n$th particle's spacetime coordinates satisfy the canonical Poisson relations $\{ x_m^\mu, P_n^\nu  \} = \delta^{\mu\nu} \delta_{mn}$. This implies that the canonical momentum $P_n^\mu$ generates translations in the spacetime coordinate $x_n^\mu$. Moreover, if it is the case that  $C_n^\pm \approx 0 $, it follows that $(P_n)_0 = \pm h_n$, which is the generator of translations in the $n$th particle's time coordinate. Said another way, $\pm h_n$ is the Hamiltonian for both the center-of-mass and internal degrees of freedom of the $n$th particle, generating an evolution of these degrees of freedom with respect to the time $x_n^0=t$, interpreted as the time measured by an inertial observer employing the coordinate system $x^\mu$.

In what follows we will employ Dirac's canonical quantization scheme~\cite{diracLecturesQuantumMechanics1964, ashtekarLecturesNonPerturbativeCanonical1991, marolfGroupAveragingRefined2002}. We promote the phase space variables of the $n$th particle to operators acting on appropriate Hilbert spaces:  $x_n^0$ and $(P_n)_0$ become canonically conjugate self-adjoint operators acting on the Hilbert space $\mathcal{H}_n^0 \simeq L^2 \left( \mathbb{R} \right)$ associated with the $n$th particle's temporal degree of freedom; $x_n^i$ and $(P_n)_i$ become canonically conjugate operators acting on the Hilbert space $\mathcal{H}_n^{\rm cm} \simeq  L^2 \left( \smash{\mathbb{R}^3} \right)$ associated with the $n$th particle's center-of-mass degrees of freedom; and $q_n$ and $P_{q_n}$ become canonically conjugate operators acting on the Hilbert space $\mathcal{H}_n^{\rm clock}$ associated with the $n$th particle's internal degrees of freedom. The Hilbert space describing the $n$th particle is thus $\mathcal{H}_n \simeq \mathcal{H}_n^{0} \otimes \mathcal{H}_n^{\rm cm} \otimes \mathcal{H}_n^{\rm clock}$. 
 
The constraint functions in Eqs.~\eqref{constraint2} and~\eqref{constraint3} become operators $C_{H_n}$ and $C_n^\pm$ acting on $\mathcal{H}_n$. The quantum analogue of the constraints is to demand that physical states of the theory are annihilated by these constraint operators
\begin{align}
C_{H_n} \kket{\Psi} = C_n^+ C_n^- \kket{\Psi} = 0 , \quad \forall \, n, \label{quantumconstraints}
\end{align}
where  $\kket{\Psi} \in \mathcal{H}_{\rm phys}$ is a physical state that is an element of the physical Hilbert space $\mathcal{H}_{\rm phys}$~\cite{rovelliQuantumGravity2004,kieferQuantumGravity2012}. The physical Hilbert space is introduced because the spectrum of $C_{H_n}$ is continuous around zero, which implies solutions to Eq.~\eqref{quantumconstraints} are not normalizable in the kinematical Hilbert space $\mathcal{K} \simeq \bigotimes_n \mathcal{H}_n$. To fully specify $\mathcal{H}_{\rm phys}$ a physical inner product must be defined, which is done in Eq.~\eqref{PhysicalInnerProduct}. Note that because $[ C_n^+ ,C_n^- ] = 0$, it follows $C_{H_n} \kket{\Psi} = 0$ if either $C_n^+ \kket{\Psi} =0$ or $C_n^- \kket{\Psi} =0$. }

\subsection*{The Page-Wootters formulation of $N$ relativistic particles}

{ \small In this subsection we recover the standard formulation of relativistic quantum mechanics with respect to a center-of-mass (coordinate) time using the Page-Wootters formalism. To do so, the physical state $\kket{\Psi}$ of $N$ particles is normalized on a spatial hypersurface by projecting a physical state $\kket{\Psi}$ onto a subspace in which the temporal degree of freedom of each particle is in an eigenstate state $\ket{t_n}$ of the operator $x_n^0$ associated with the eigenvalue $t \in \mathbb{R}$ in the spectrum of $x_n^0$, $x_n^0\ket{t_n} = t \ket{t_n}$. Explicitly, 
\begin{align}
\Pi_t \otimes I_S \kket{\Psi} =  \ket{t} \ket{\psi_S(t)},
\label{SpatialHypersurfaceProjection}
\end{align}
where $I_S$ denotes the identity on $\mathcal{H}_S \simeq  \bigotimes_n\mathcal{H}_n^{\rm cm} \otimes \mathcal{H}_n^{\rm clock}$, $\Pi_t \ce \ket{t}\!\bra{t}$ is a projector onto the subspace of $\mathcal{H}$ in which the temporal degree of freedom of each particle is in a definite temporal state $\ket{t}\ce \bigotimes_n \ket{t_n}$ associated with the eigenvalue $t$. Equation~\eqref{SpatialHypersurfaceProjection} defines the conditional state
\begin{align}
\ket{\psi_S(t)} &\ce   \bra{t} \otimes I_S  \kket{\Psi} \in  \mathcal{H}_S,\label{defconditinalstate}
\end{align}
which describes the state of the center-of-mass and internal degrees of freedom of $N$ particles conditioned on their temporal degree of freedom being in the state $\ket{t_n}$. We demand that this state is normalized $\braket{\psi_S(t) | \psi_S(t)}=1$ for all $t\in\mathbb{R}$. This implies that the physical states are normalized with respect to the inner product~\cite{smithQuantizingTimeInteracting2017}
\begin{align}
\braket{\braket{\Psi | \Psi }}_{\rm PW} &\ce \braket{\braket{\Psi | \Pi_t \otimes I_S | \Psi }}  = \braket{\psi_S(t) |  \psi_S(t) } = 1,
\label{PhysicalInnerProduct}
\end{align}
for all $t \in \mathbb{R}$.

Note that the set $\left\{ \Pi_t , \, \forall \, t \in \mathbb{R} \right\}$ constitutes a projective valued measure (PVM) on the Hilbert space $\bigotimes_n \mathcal{H}_n^0$, since $\braket{t|t'} = \delta(t-t')$ and $\int dt \, \Pi_t = I^0$, where $I^0$ is the identity on $ \bigotimes_n \mathcal{H}_n^0$. Given this observation and the definition of the conditional state in Eq.~\eqref{defconditinalstate}, it is seen that the physical state $\kket{\Psi}$ is  entangled relative to $\mathcal{H}_n^{\rm cm} \otimes \mathcal{H}_n^{\rm clock}$,
\begin{align}
\kket{\Psi} &=  \left(\int dt \, \Pi_t  \otimes I_S  \right)\kket{\Psi} =  \int dt\, \ket{t}\ket{\psi_S(t)}.
\label{entangledPhysicalState}
\end{align}
We emphasize that this entanglement is with respect to a partitioning of the kinematical Hilbert space $\mathcal{K}$, and thus is not physical (i.e., not gauge invariant)~\cite{hoehn2020}.

We consider physical states that satisfy 
\begin{align}
C_n^+ \kket{\Psi} = \left[ (P_n)_0 + h_n \right] \kket{\Psi} = 0,
\end{align}
for all $n$, where $h_n$ is the operator equivalent of Eq.~\eqref{defofh}. This amounts to demanding that the conditional state of the system has positive energy as measured by $h_n$.

We now show that the conditional state $\ket{\psi_S(t)}$, defined in Eq.~\eqref{defconditinalstate}, satisfies the Schr\"{o}dinger equation in the parameter $t$. Recall that $[x_n^0,(P_n)_0] = i $, and hence the operators $P_n^0$ generate translations in~$x_n^0$,
\begin{align}
\ket{t'_n}  = e^{-i (t' - t)  (P_n)_0} \ket{t_n}, \label{generator}
\end{align}
where $\ket{t_n}$ and $\ket{t_n'}$ are eigenkets of the operator $x_n^0$ with respective eigenvalues $t$ and $t'$. Now consider how $\ket{\psi_S(t)}$ changes with the parameter~$t$:
\begin{align}
i \frac{d}{dt} \ket{\psi_S(t)} &= i \frac{d}{dt}   \left( \bigotimes_n \bra{t_n} \otimes I_S\right) \kket{\Psi} \nn \\
&=   \left( \sum_m \bra{t} (P_m)_0 \otimes I_{\neg m }^0 \otimes I_S   \right)  \kket{\Psi}, \label{derivation1}
\end{align}
where $I_{\neg m }^0$ denotes the identity operator on all of the Hilbert spaces $\mathcal{H}_n^0$ for which $n\neq m$ and the derivative with respect to $t$ was evaluated using Eq.~\eqref{generator}. The constraint $C_n^+  \kket{\Psi} =0$ can be rewritten as
\begin{align}
(P_m)_0 \otimes I_{\neg m }^0 \otimes I_S   \kket{\Psi}
= - I^0 \otimes h_m \otimes I_{S_{\neg m}}  \kket{\Psi}, \label{constraintrearanged}
\end{align}
for all $m$ where $h_m$ is the  operator equivalent of Eq.~\eqref{defofh} acting on $ \mathcal{H}_m^{\rm cm} \otimes \mathcal{H}_m^{\rm clock}$ and $I_{S_{\neg m}}$ is the identity on \mbox{$\bigotimes_{n\neq m} \mathcal{H}_m^{\rm cm} \otimes \mathcal{H}_m^{\rm clock}$}. Substituting Eq.~\eqref{constraintrearanged} into Eq.~\eqref{derivation1} yields
\begin{align}
i \frac{d}{dt} \ket{\psi_S(t)}
&=  \left(\sum_m   \bra{t}  \otimes h_m  \otimes I_{S_{\neg m}}  \right)    \kket{\Psi} \nn \\
&=  \sum_m     \int dt' \,   \braket{t | t' }   h_m  \otimes I_{S_{\neg m}}   \ket{\psi_S(t')} \nn \\
&=  \sum_m     h_m  \otimes I_{S_{\neg m}}   \ket{\psi_S(t)},
\label{derivation2}
\end{align}
where the second equality is obtained using Eq.~\eqref{entangledPhysicalState}. Equation \eqref{derivation2} asserts that the conditional state satisfies the Schr\"{o}dinger equation
\begin{align}
i \frac{d}{dt} \ket{\psi_S(t)} &= H_S \ket{\psi_S(t)}, \label{SchrodingerEq}
\end{align}
where $H_S \ce \sum_m     h_m  \otimes I_{S_{\neg m}}$ is the relativistic of all $N$ particles.}

\subsection*{Leading relativistic expansion of the conditional time probability distribution}
\label{DerivationOfCondtionalProb}
{\small The Hamiltonian in Eq.~\eqref{defofh} can be expressed as
\begin{align}
h_n = H_n^{\rm clock} + H_n^{\rm cm } + H_n^{\rm int},
\end{align}
where we have specialized to Minkowski space, dropped an overall constant $mc^2$, and defined the center-of-mass Hamiltonian $H_n^{\rm cm} \ce \mathbf{P}^2_n/2m$ and the leading order relativistic contribution
\begin{align}
H_n^{\rm int} &\ce 
- \frac{1}{mc^2} \left( H_n^{\rm cm}\otimes  H_{n}^{\rm clock}   + \frac{1}{2} \left(H_n^{\rm cm}\right)^2  \right),
\label{interactionHamiltonian}
\end{align}
which is derived by expanding Eq.~\eqref{defofh} in both $\mathbf{P}_n/mc $ and $H_{n}^{\rm clock}/mc^2$. }

{\small  Let us define the free evolution of the center-of-mass and internal clock degrees of freedom respectively as
\begin{equation}
\bar{\rho}^{\rm cm}_n(t) \ce e^{-iH_n^{\rm cm} t} \rho_n^{\rm cm} e^{iH_n^{\rm cm} t},
\end{equation}
and
\begin{equation}
 \bar{\rho}_n(t) \ce e^{-iH_n^{\rm clock} t} \rho_n e^{iH_n^{\rm clock} t},
\end{equation}
for $n \in \{A,B\}$. Then the reduced state of the clock to leading relativistic order is
\begin{align}
\rho_n(t) &= \tr_{\rm cm} \left(e^{-i H_n^{\rm int} t}   \bar{\rho}_n^{\rm cm}(t) \otimes \bar{\rho}_n(t)  e^{i H_n^{\rm int} t} \right) \nn \\
&= \bar{\rho}_n(t)   + it \frac{\braket{H_n^{\rm cm }}}{mc^2} \left[  H_n^{\rm clock}   , \bar{\rho}_n(t)   \right].  
\label{expansion}
\end{align}
Using Eq.~\eqref{expansion} the integrands defining the conditional probability distribution in Eq.~\eqref{ConditionalProbabilityDis} may be evaluated perturbatively
\begin{align}
\tr \big( E_{n}(\tau_n)   \rho_{n}(t) \big) &= \tr \big( E_{n}(\tau_n)   \bar{\rho}_n(t)\big)
\nn \\
&\quad + it \frac{\braket{H_n^{\rm cm }}}{m c^2}   \tr \big(  E_{n}(\tau_n)   \left[ H_{n}^{\rm clock} , \bar{\rho}_n(t)   \right]\big).
\label{integrand1}
\end{align}

Suppose that the fiducial state $\ket{\psi^{\rm clock}_n} \in \mathcal{H}_n^{\rm clock}$ of the clock is Gaussian with a spread $\sigma$,  then the first term in Eq.~\eqref{integrand1} is
\begin{align}
\tr \left [ E_{n}(\tau_n)   \bar{\rho}_n(t)\right] &=
\abs{ \bra{\tau_n} e^{- i H_n^{\rm clock} t }  \ket{\psi^{\rm clock}_n} }^2 \nn \\
&=  \abs{\int_{\mathbb{R}} d \tau' \, \braket{\tau_n |\tau_n'}  \frac{e^{-\frac{(\tau'-t)^2}{2 \sigma^2}}  }{\pi^{\frac{1}{4}} \sqrt{\sigma}}   }^2 \nn \\
&= \frac{e^{-\frac{(\tau-t)^2}{\sigma^2}}}{\sqrt{\pi} \sigma} ,
\label{FirstTerm}
\end{align}
where we used the orthogonality of the clock states, $\braket{\tau_n| \tau_n'} = \delta(\tau-\tau')$, which holds for an ideal clock.

Defining $\ket{\psi^{\rm clock}_n(t)} \ce e^{- i H_n^{\rm clock} t }  \ket{\psi^{\rm clock}_n}$, the trace in the second term of Eq.~\eqref{integrand1} is 
\begin{align}
&\tr \left [ E_{T_n}(\tau_n)   \left[ H_{n}^{\rm clock} , \bar{\rho}_n(t)   \right]\right]\nn \\
&= \bra{\tau_n} \left[ H_{n}^{\rm clock} , \bar{\rho}_n(t)   \right]\ket{\tau_n} \nn \\
 &= \bra{\tau_n }   H_n^{\rm clock} \ket{\psi^{\rm clock}_n(t)}  \braket{\psi^{\rm clock}_n(t) | \tau_n}  \nn \\
 &\quad 
-\braket{\tau_n | \psi^{\rm clock}_n(t)}  \bra{\psi^{\rm clock}_n(t)}   H_n^{\rm clock}   \ket{\tau_n} \nn \\
&=  \big( \bra{\tau_n }   H_n^{\rm clock} \ket{\psi^{\rm clock}_n(t)}  - \bra{\psi^{\rm clock}_n(t)}    H_n^{\rm clock}   \ket{\tau_n}\big) \nn \\
&\quad \times \frac{ e^{-\frac{(\tau-t)^2}{2 \sigma^2}}}{\pi ^{\frac{1}{4}} \sqrt{\sigma}}.
\label{trace2ndTerm}
\end{align}
It follows from the  covariance relation in Eq.~\eqref{CovarianceCondition}  that the clock states satisfy
\begin{align}
\ket{(\tau + \tau')_n} = e^{-iH^{\rm clock}_n \tau'}\ket{\tau_n},
\end{align}
which implies that $H^{\rm clock}_n\equiv -i \partial/ \partial\tau $ is the displacement operator in the $\ket{\tau_n}$ representation~\cite{wisemanQuantumMeasurementControl2010}. This observation allows us to evaluate the probability amplitudes in Eq.~\eqref{trace2ndTerm}
\begin{align}
\bra{\tau_n }   H_n^{\rm clock} \ket{\psi^{\rm clock}_n(t)} &= - i  \frac{\partial}{\partial  \tau}   \frac{e^{-\frac{(\tau-t)^2}{2 \sigma^2}} }{\pi^{\frac{1}{4}} \sqrt{\sigma}} \nn \\
&=  i\frac{e^{-\frac{(\tau-t)^2}{ 2 \sigma^2}}}{\pi^{\frac{1}{4}} \sqrt{\sigma}}   \frac{\tau-t}{\sigma^2},
\label{appendixSimp1}
\end{align}
which simplifies Eq.~\eqref{trace2ndTerm} to
\begin{equation}
\tr \left( E_{n}(\tau_n)   \left[ H_{n}^{\rm clock} , \bar{\rho}_n(t)   \right]\right) 
=
2i\frac{e^{-\frac{(\tau-t)^2}{ \sigma^2}}  }{\sqrt{\pi}  \sigma}    \frac{\tau-t}{\sigma^2},
\end{equation}
and together with Eq.~\eqref{FirstTerm}, Eq.~\eqref{integrand1} reduces to
\begin{align}
\tr \left [ E_{n}(\tau_n)   \rho_{n}(t) \right] &= \tr \left [ E_{n}(\tau_n)   \bar{\rho}_n(t)\right] \nn \\
&\quad 
+ it \frac{\braket{H_n^{\rm cm }}}{mc^2}  \tr \left [ E_{n}(\tau_n)   \left[ H_{n}^{\rm clock} , \bar{\rho}_n(t)   \right]\right]   \nn \\
&= \frac{e^{-\frac{(\tau-t)^2}{\sigma^2}}}{\sqrt{\pi} \sigma}  \nn \\
&\quad 
+ it \frac{ \braket{H_n^{\rm cm }} }{mc^2}\left(2i\frac{e^{-\frac{(\tau-t)^2}{ \sigma^2}}  }{\sqrt{\pi}  \sigma}    \frac{\tau-t}{\sigma^2}\right)  \nn \\
&= \frac{ e^{-\frac{(\tau-t)^2}{\sigma^2}}}{\sqrt{\pi} \sigma^2} \left[1 
-  2   \frac{ \braket{H_n^{\rm cm }}  }{mc^2} \frac{t (  \tau - t) }{ \sigma^2}   \right].
\label{integrand2}
\end{align}
Using Eq.~\eqref{integrand2} the conditional probability defined in Eq.~\eqref{ConditionalProbabilityDis} can be evaluated, yielding Eq.~\eqref{ConditionalProbabilityDistirbutionMinkowski}
\begin{align}
&\prob \left[T_A = \tau_A \ | \ T_B = \tau_B \right] \nn \\
&= \frac{\int dt \, \tr \left [ E_A(\tau_A)   \rho_{A}(t) \right]  \tr \left [ E_B(\tau_B)   \rho_{B}(t) \right]}{\int dt \, \tr \left [ E_B(\tau_B)   \rho_{B}(t) \right]} \nn \\
&= \frac{e^{- \frac{(\tau_A-\tau_B)^2}{2 \sigma^2}}}{\sqrt{2 \pi} \sigma} \frac{1 +   \frac{ \braket{H^{\rm cm}_A} + \braket{H^{\rm cm}_B}}{2 mc^2}   - \frac{\braket{H^{\rm cm}_A} - \braket{H^{\rm cm}_B} }{2 mc^2}  \frac{\tau_A^2 - \tau_B^2}{\sigma^2}     }{1 + \frac{\braket{H^{\rm cm}_B}}{mc^2} } \nn \\
&= \frac{e^{- \frac{(\tau_A-\tau_B)^2}{2 \sigma^2}}}{\sqrt{2 \pi} \sigma}  \left( 1 +   \frac{\braket{H^{\rm cm}_{A}}- \braket{H^{\rm cm}_{B}}}{mc^2}  \frac{\sigma^2 - \tau_A^2 + \tau_B^2}{2 \sigma^2}  \right).
\label{endResult}
\end{align}

Instead, for a two-level atom as a clock we would have had $\mathcal{H}^{\rm clock} \simeq \mathbb{C}^2 = \Span \{ \ket{0}, \ket{1}\}$, $H^{\rm clock} = \Omega \sigma_z$, and the covariant time observable with respect to the group generated by $\Omega \sigma_z$, i.e., $\{ E(\tau) = \ket{\tau}\!\bra{\tau}, \ \forall \tau \in (0, 2 \pi / \Omega]\}$, where $\ket{\tau} = \frac{1}{\sqrt{2}} \left( \ket{0} + e^{2 i \Omega \tau} \ket{1} \right)$. For such clock states, we have $ \braket{\tau |\tau'} = \tfrac{1}{2}[1 + e^{2 i \Omega (\tau'-\tau)} ]$, leading to a modification of the last equality in Eq.~\eqref{FirstTerm} and the results that follow. Nonetheless, a similar analysis should lead to an analogous quantum time dilation effect that will be modified by the specific details of the clock. The details of clocks described by discrete spectrum Hamiltonians and the associated covariant time observables have recently been discussed in a related context~\cite{hoehn2020}.} \\

\section*{ Data availability statement}

{ \small Data sharing not applicable to this article as no datasets were generated or analysed.}

\bibliographystyle{nature}
\bibliography{ProbabilisticTimeDilation}

\noindent{\bf Acknowledgments}

 This work was supported by the Natural Sciences and Engineering Research Council of Canada and the Dartmouth Society of Fellows. We'd like to thank Philipp A. H\"{o}hn and Maximilian P. E. Lock for useful discussions.

\noindent{\bf Author contributions}

A.R.H.S. and M.A. jointly conceived of the ideas presented and contributed to writing this article. 

\noindent {\bf Competing interests:} The authors declare no competing interests.

\onecolumngrid

\title{Supplementary Information: Quantum clocks observe classical and quantum time dilation 
}

\author{Alexander R. H. Smith}
\affiliation{Department of Physics, Saint Anselm College, Manchester, New Hampshire 03102, USA} 
\affiliation{Department of Physics and Astronomy, Dartmouth College, Hanover, New Hampshire 03755, USA}

\author{Mehdi Ahmadi}

\affiliation{Department of Mathematics and Computer Science, Santa Clara University, Santa Clara, California 95053, USA}

\date{\today}

\maketitle

\onecolumngrid

\section*{Supplementary Note 1: Recovering the Klein-Gordon equation in the Page-Woottters formalism}

For simplicity, let us consider a single particle situated in Minkowski space, so that the constraint in  Eq.~(31) of the main text becomes
\begin{align}
C_{H} \kket{\Psi}  = \left( \eta^{\mu\nu}P_\mu P_\nu \otimes I_C + I_0 \otimes I_{\rm cm} \otimes \left(m + H_{\rm clock}/c^2\right)^2c^2 \right) \kket{\Psi}= 0, \label{constraintAppendix}
\end{align}
where $\eta_{\mu \nu}$ denotes the Minkowski metric and we have suppressed the subscript $n$. Given a physical state satisfying this constraint, Eq.~(36) of the main text defines the conditional state of the center-of-mass and internal degrees of freedom of the particle 
\begin{align*}
\ket{\psi_S(t, \mathbf{x})} \ce  \big( \bra{t} \bra{\mathbf{x}} \otimes I_{C} \big)  \kket{\Psi} ,
\end{align*}
where $x^0 \ket{t} = t \ket{t}$ and $\ket{\mathbf{x}} \ce \ket{x^1}\ket{x^2}\ket{x^3}$ with $\ket{x^i}$ denoting an eigenstate of the operator $x^i$. Now consider the action of the d'Alambertian  operator $\Box \ce \eta^{\mu \nu} \partial_\mu \partial_\nu$ on the conditional state
\begin{align}
\Box\ket{\psi_S(t, \mathbf{x})}&= \eta^{\mu \nu}\partial_\mu \partial_\nu   \big( \bra{t} \bra{\mathbf{x}} \otimes I_{C} \big)  \kket{\Psi}  \nn \\
&=  \eta^{\mu \nu}\partial_\mu \partial_\nu   \left( \bra{t_0} e^{iP_0 t}\bra{\mathbf{x}_0} e^{i P_i x^i } \otimes I_{C} \right)  \kket{\Psi}  \nn \\
&=  \big( \bra{t} \bra{\mathbf{x}} \otimes I_{C} \big) \big( \eta^{\mu \nu} P_\mu P_\nu \otimes I_{C} \big)  \kket{\Psi}  \nn \\
&=  - \big( \bra{t} \bra{\mathbf{x}} \otimes I_{C} \big) \left[ I_0 \otimes I_{\rm cm} \otimes \left(m + H_{\rm clock}/c^2\right)^2c^2  \right]  \kket{\Psi}  \nn \\
&=  -  I_0 \otimes I_{\rm cm} \otimes \left(m + H_{\rm clock}/c^2\right)^2c^2     \ket{\psi_S(t, \mathbf{x})},\label{stepsToKG}
\end{align}
where the third equality is obtained using Eq.~\eqref{constraintAppendix}. Upon rearranging Eq.~\eqref{stepsToKG} we find that the conditional state satisfies 
\begin{align}
\left[ \Box + \left(m + H_{\rm clock}/c^2\right)^2c^2 \right] \ket{\psi_S(t, \mathbf{x})} = 0,
\label{KleinGordonEq}
\end{align}
where we have suppressed the identity operators $I_0$, $I_C$, and $I_{\rm cm}$. If one supposes $H_{\rm clock}$ vanishes, then Eq.~\eqref{KleinGordonEq} reduces to the usual Klein-Gordon equation.

\section*{Supplementary Note 2: Justification for using the Page-Wootters formalism}

One might question why a more standard formulation of relativistic quantum mechanics was not used. We feel the following excerpt, that has been edited for clarity, from Feynman's 1964 Messenger lectures delivered at Cornell University justifies why one should adopt a plurality of theoretical approaches to describe a given phenomena:
\begin{quote}
\quad ``Consider two identical theories $A$ and $B$, which look completely different psychologically and have different ideas in them, but all their consequences are exactly the same. A thing that people often say is how are we going to decide which one is right?

\quad No way! Not by science because both theory $A$ and theory $B$ agree with experiment to the same extent so there is no way to distinguish one from the other. So if two theories, though they may have deeply very different ideas behind them, can be shown to be mathematically equivalent then people usually say in science that the theories can not be distinguished.  

\quad However, theories $A$ and $B$ for psychological reasons, in order to guess new  theories, are very far from equivalent because one gives the scientist very different ideas than the other. By putting a theory in a given framework you get an idea of what to change. It may be the case that a simple change in theory~$A$ may be a very complicated change in theory~$B$. In other words, although theories $A$ and $B$ are identical before they're changed, there are certain ways of changing one that look natural which don't look natural in the other. 

\quad Therefore, psychologically we must keep all the theories in our head and every theoretical physicist  that is any good knows six or seven different theoretical representations for exactly the same physics, and knows  that they are all equivalent, and that nobody is every going to be able to decide which one is right at that level. But they keep these representations in their head hoping they will give them different ideas for guessing.''
\end{quote}
In this case, the Page-Wootters formalism suggested to formulate time dilation in terms of the conditional probability distribution in Eq.~(12) of the main text.

\section*{Supplementary Note 3: Proof of desiderata of physical clocks theorem}

The theorem stated in the Results is a summary of well-known results of quantum parameter estimation~\cite{holevoProbabilisticStatisticalAspects1982,braunsteinStatisticalDistanceGeometry1994,braunsteinGeneralizedUncertaintyRelations1996,wisemanQuantumMeasurementControl2010}. We summarize here how the two properties of the theorem follow from the covariance properties of the POVM. 

The first statement follows from a direct computation of the average of $T_{\rm clock}$ on the state~$\rho(\tau)$
\begin{align}
\braket{T_{\rm clock}}_{\rho(\tau)} &=  \int_{G} d\tau' \, \tr \left[ E(\tau') \rho(\tau) \right] \tau' \nn \\
&=  \int_{G} d\tau' \, \tr \left[ U^\dagger_{\rm clock}(\tau) E(\tau') U_{\rm clock}(\tau) \rho \right] \tau' \nn \\
&=  \int_{G} d\tau' \, \tr \left[ E(\tau' - \tau)  \rho\right] \tau' \nn \\
&=  \int_{G} d\tau' \, \tr \left[ E(\tau') \rho \right] \left(\tau' + \tau\right) \nn \\
&=  \tau, \nn 
\end{align}
where the third equality follows from Eq.~(3) of the main text, the fourth equality follows from a change of variables $\tau' \to \tau'-\tau$, and in arriving at the last equality we used the fact that by construction $\braket{T_{\rm clock}}_{\rho} = 0$.

The second statement follows in a similar manner
\begin{align}
{\braket{(\Delta T_{\rm clock})^2}_{\rho(\tau)}} &=  \int_{G} d\tau' \, \tr \left[ E(\tau') \rho(\tau) \right] \tau'^2 - \braket{T_{\rm clock}}_{\rho(\tau)}^2 \nn \\
&=  \int_{G} d\tau' \, \tr \left[ E(\tau'-\tau) \rho\right] \tau'^2 - \tau^2 \nn \\
&=  \int_{G} d\tau' \, \tr \left[ E(\tau') \rho\right] \left(\tau'+\tau\right)^2 - \tau^2 \nn \\
&= {\braket{(\Delta T_{\rm clock})^2}_{\rho}}. \nn
\label{VarainceInTime}
\end{align}

\end{document}